# Laboratories for Data Communications and Computer Networks[*][†]

*Rohit Goyal, Steve Lai, Raj Jain, Arian Durresi*
*Department of Computer and Information Science*
*The Ohio State University*
*395 Dreese Lab, 2015 Neil Avenue*
*Columbus, OH 43210-1277, U.S.A.*
*E-mail: goyal@cis.ohio-state.edu*

**Abstract** – *In this paper we describe a hands-on laboratory oriented learning and instructional package that we have developed for data communications and networking. The package consists of a software tool, together with instructional material for a laboratory based networking curriculum. The software is based on a simulation environment that enables the student to experiment with various networking protocols, on an easy to use graphical user interface (GUI). Data message flows, packet losses, control/routing message flows, virtual circuit setups, link failures, bit errors etc., are some of the features that can be visualized in this environment. The student can also modify the networking components provided, as well as add new components using the C programming language.*

*The instructional material consists of a set of laboratory exercises for flow and error control (HDLC), IEEE 802.3 CSMA/CD protocol, the token ring protocol, interconnecting LANs via bridges, TCP congestion avoidance and control, IP fragmentation and reassembly, ATM PNNI routing and ATM policing, all using the simulation package. The laboratory exercises have facilitated the development of a networking curriculum based on both the traditional computer networking principles, as well as the new technologies in telecommunication networking.*

*The laboratory environment has been used in the networking curruculum at The Ohio State University, and is being piloted at other universities. The entire package is freely available over the Internet.*

## 1 Introduction

The field of computer networking is entering the exponential growth phase of its life cycle and is becoming an important part of every computer science curriculum. The plans for the National Information Infrastructure - our national information highway - have helped fuel the demand for people with a networking education. The enrollment in networking courses at universities is growing exponentially just like the number of networks that are being connected to the Internet.

The field of computer networking is also changing rapidly. With the recent development of ATM networks and plans for their rapid deployment, most courses in networking are fast becoming out-of-date. There is a need for networking courses to integrate the concepts from traditional computer networking with those of the telecommunication networks.

Laboratory work is an indispensable component of any engineering or applied science curriculum. This applies to networking courses as well. Unfortunately, most computer networking courses do not have laboratory components. Even if there is a lab, it is usually related to the higher layer protocols, such as electronic mail, file transfer, and remote login. Very few departments have facilities for students to experiment with lower layer protocols, such as data link layer and network layer. Such environments require a large number of computers interconnected via various network technologies. Their use by a relatively small group of students is economically unjustifiable. Even if a department could provide one type of network, e.g., Ethernet, it would be difficult to provide a wide variety of networks, such as token ring, FDDI, DQDB, ATM, etc. The only practical way to cover such a variety is to use simulated networks whose protocol components can be easily interchanged or modified by the student. This can be done by providing a learning environment in which all the networking modules except one are available. The student is expected to design and implement the missing module and experiment with it.

In this paper, we present a laboratory-oriented hands-on curriculum for data communications and computer networking. We have developed software that allows students to experiment with various computer networking, data communications and telecommunications concepts. Students, sitting on their workstations can exercise various options available to network nodes and visually see the impact of their actions. Data message flows, packet losses, control/routing message flows, virtual circuit setups, link failures, bit errors, etc. can be seen by the students at a slow (visible) speed.

[*]This work is supported by the National Science Foundation. Grant Number CDA-9522319, 7/1/95–6/30/98

[†]The software package developed in this project is available from http://www.netlab.ohio-state.edu/cise

These lab exercises also facilitate the instruction of the applications of protocol engineering principles. At present, protocol engineering is accessible only to graduate students because the subject is taught in a theoretical fashion. Use of hands-on exercises allows students to apply the protocol specification, validation, and testing techniques to their protocol designs.

## 2 The Laboratory Environment

The laboratory package consists of a software tool, together with instructional material for a laboratory based networking curriculum. The key features of this package are:

- A hands-on laboratory-oriented curriculum requiring students to develop protocol modules and experiment with their designs.

- Application of protocol engineering principles to protocol development.

- Integration of traditional data communication, computer networking, and telecommunications concepts.

- Use of a simulation environment to allow experimentation with "what if" scenarios.

- Use of multimedia and visual techniques to illustrate the functioning of various networking protocols.

- Availability of the source code for the package, portability of the environment across a variety of platforms, and easy dissemination over the Web.

The software is a simulation environment that abstracts the essential content of various networking protocols and makes it available for experimentation. In particular, the software enables the student to experiment with various networking protocols, on an easy to use graphical user interface (GUI). Students can visually, as well as quantitatively, study protocol correctness and performance by changing network topologies and parameter values. The simulation package allows easy accessibility to modify the networking components provided, as well as the ability to add new components using the C programming language.

Using a simulation environment has several advantages over experiments using a real network. Building a simulation tool is much cheaper than buying real equipment. The simulation environment allows the student to experiment with, and modify existing protocols, as well as develop and test new protocols. The complete instructional package can be easily and widely disseminated through the world wide web.

### 2.1 The Network Simulator

The core of the simulation package is based on Netsim, a widely used event-driven simulator for networking protocols. The simulator has been completely written is C, and its entire source is publicly available for use and modification. In the simulator, a network consists of components connected to one another. Examples of components are protocol layer modules, links, network nodes etc. Components communicate with one another via events. A global event heap stores events in the order in which they are supposed to occur. Events in the event heap are associated with a particular component, and each component has an event handler function that performs the necessary tasks on the occurrence of an event. Examples of events are sending and receiving of packets from one component to another, expiration of timers within a component etc.

Only a knowledge of C programming is needed by the student and the instructor for using and grading the lab exercises provided in the package. An understanding of event driven programming, and the structure of the simulator model is needed only for adding new components, or developing new lab exercises. In any case, the entire source code for the simulator is written in C.

### 2.2 The Graphical User Interface

The simulator has an easy to use GUI that allows the student to graphically experiment with the network under study. The GUI is written using the Tcl/Tk package to ensure compatibility across a wide variety of platforms. The simulator interface consists of a main window that contains the network of nodes connected by links and interconnection devices. Figure 1 illustrates a network of Ethernet (10Base5) LANs connected together by bridges. Figure 2 shows an end-to-end TCP connection separated by a network cloud. In this case, each TCP end system is represented by four nodes in the simulator window. The nodes correspond to four layers in the end-system – physical, datalink, transport and application. This depicts a simple abstraction of an end system with the essential network functionality needed to study the TCP protocol. The top of the main simulator window has a menu bar that enables the user to control the simulation runs. The menu bar contains functionalities that allow the user to load the network configuration files, start and stop simulations, set delays etc. The complete contents of the menu bar are listed in table 1.

At the bottom of the main simulator window is the status bar that displays current information about the simulation under study. Table 2 lists the contents of the status bar. In addition to the main window, there may be several other sub windows depending on the experiment being performed. For example, in figure 1, the three sub-windows contain the real-time filtering databases of each of the three bridges respectively. Each database is in the form of a table that shows the current routes learnt by the bridge. The entries in the table change as the information in the bridge is updated. Figure 2 shows an experiment with the TCP protocol. The network consists of a two TCP nodes separated by a network. Each node has four protocol layers – physical, datalink, transport and application. For each application layer, there is a send/receive window where the user can load a graphic and send it to the peer layer. Each TCP layer has a congestion window plot and a sequence number plot associated with it. The congestion window plot shows the variation of the congestion window with time. The sequence number window has 3 plots associated with it:

- The sequence number of the first byte of each segment sent by the TCP (red dots).

- The ACK number in the acknowledgement received

Table 1: *Main Window Menu Bar Contents*

| Menu | Submenu | Description |
|---|---|---|
| File | Load | Load configuration file |
|  | Exit | Stop the program and exit |
| Edit | Raise | Raise the sub-windows over the main window |
| Run | Run | Start the simulation |
|  | Single Step | Enters single step mode using the space bar for a step. |
|  | Pause | Pause the simulation. |
|  | Resume | Resume the paused simulation. |
|  | Delay | Set delay between events. This determines the speed of packets seen in the window. |
|  | Stop Time | Change the simulation stop time. |
|  | Inc/Dec Debug Level | Increament/Decrement Debug Level, which is used in dprintf(). |
| Help | About | About the CISE Project. |
|  | How to use | Help text. |

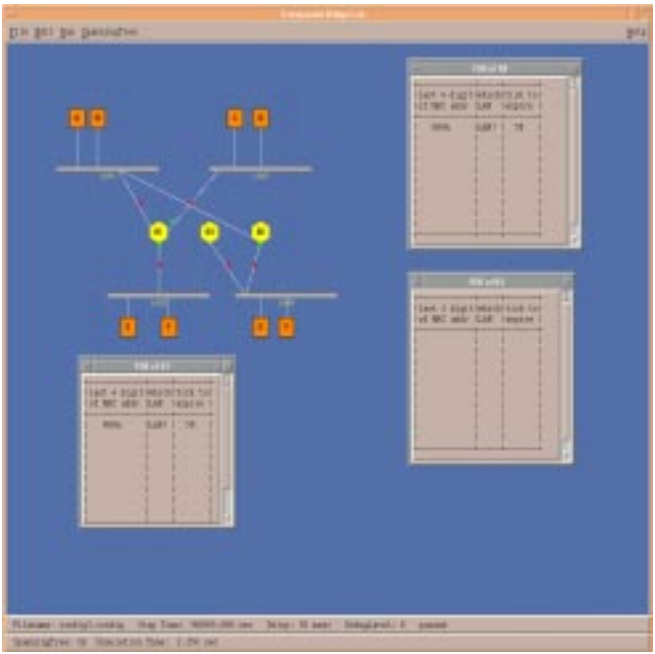

Figure 1: *Simulator GUI for Bridges*

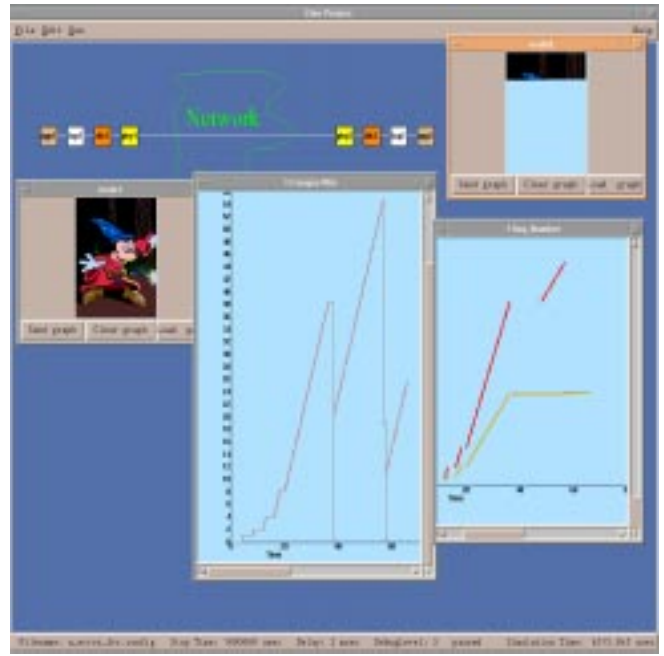

Figure 2: *Simulator GUI for TCP*

Table 2: *Main Window Status Bar Contents*

| Field | Description |
| --- | --- |
| Filename | The configuration file name. |
| Stop Time | The end time for the simulation |
| Delay | Delay between events. |
| DebugLevel | Current debug level used in dprintf(). |
| Simulation time | The clock in the simulator. |

by the sender (gold dots).

- The sequence number of the packets received by the receiver (dark brown dots).

The GUI also provides the flow of packets within the network and space-time diagrams to show the acknowledgement behavior of packets. These packets are often color coded based on their types and the network states. In this way, students can not only see the flow of packets through the protocol layers, and across the network, they can also look at the corresponding dynamic behavior of the protocol parameters within each network module.

In addition, the package contains utilities, that facilitate the submission of graphical results on paper. This allows the student to submit postscript based graphs together with text output of statistics, for evaluation of the assignment.

## 2.3 Instructional Material

The instructional material consists of a set of extensive laboratory exercises using the software package. The main goal of these exercises is to provide a hands-on experience to the fundamentals of network protocol design, implementation and performance. A typical exercise consists of an experimental study of the performance of a given protocol, followed by an enhancement or a modification of the protocol. For example, we have developed exercises for flow and error control (HDLC), IEEE 802.3 CSMA/CD protocol, the token ring protocol, interconnecting LANs via bridges, TCP congestion avoidance and control, IP fragmentation and reassembly, ATM PNNI routing and ATM policing, all using our simulation package. These labs are further discussed in section 4.

## 3 Lab Methodology

In order to facilitate the experimentation of different networking protocols and concepts, we provide a programming environment in which all the networking modules are available, except for the one under study. Students design and implement the missing module and experiment with it. The experiments are based on various topologies and parameters. These experiments are designed to provide an in depth understanding of the various features of the protocol/algorithm under study.

Each lab assignemt consists of several stages.

- The students first experiment with a *demo* version of the module or layer under study. In this way they can understand the features of the module. The *demo* version serves also as a model of behavior and the performance for their module.

- Students are asked to answer several questions based on designed experiments with the *demo*. These questions range from observation of basic events during the course of the simulation, to qualitative and quantitative analysis of the performance of the protocol.

- Students are given the outline of the module to be designed, and some programming primitives for designing the module. They develop this module using the C programming language. They experiment with their module layer using different configurations and parameters.

- Statistics are produced in graphical and/or tabular format for submission and grading.

The following steps are used by the student to run an experiment. The TCP experiment is used as an example.

- First, load a configuration file. This can be done by File/Load or by appending the config file at the command line of this program.

- To send a graph, "Load Graph" first and then "Send Graph." and the simulator will load the graph and start running. Packes will be seen travelling between the end stations. These packets are color coded based on regular packets, acknowledgments, corrupted packets, retransmitted packets.

- Change the delay to make it run faster/slower. Use "Run/Single step" and space bar to see it step by step.

- Change Debug Level (from 0–3) and use `dprintf(int debug_level, ``format'', variables)` in the code to print out debug information. `dprintf` allows the student to insert debugging statements in various levels of their code. The value of the debug level determines which `dprintf` statements will be printed.

# 4 Description of the Labs Developed

The following nine labs have been developed so far using this simulation package.:

1. *Interlayer Communication:* In this lab, students are introduced to the basic concepts of layered network architectures. They learn how to use primitives in interlayer communications. This lab also serves as an introduction to the programming environment and graphical interface used in the other labs.

2. *Sliding Window Protocol and Go-Back-N ARQ:* In this lab, students develop a simple datalink layer module that performs flow control using the sliding window protocol, and error control using the Go-Back-N mechanism. The students then experiment withe these protocols for various topologies and link error rates, and assess their performance.

3. *IEEE 802.3 Media Access Control:* In this lab, students experiment with the main ideas of CSMA/CD Media Access Control (MAC) protocol, including collision detection, jam pattern transmission and exponential backoff before retransmission. They implement the collision and exponential backoff algorithms. The graphical interface helps in visualising the transmission of frames on a shared media, occurrence of collision, and retransmission.

4. *IEEE 802.5 Token Ring:* This lab focuses on three features of the Token Ring protocol – One bit delay, the priority stack, and simple token management. First, the students experiment with a given token ring module, and understand the basic mechanisms that control token access. Then, they implement the key parts of the above features of the token ring protocol.

5. *Transparent Bridges:* This lab explains the following features of bridges.

   - The promiscuous listen, and the store and forward capabilities which permits stations designed to operate on a single LAN, to work in a multi-LAN environment.
   - The learning techniques used by bridges to determine where stations and LANs are located.
   - The spanning tree algorithm used by a bridge to dynamically discover a fully connected, loop free subset of the network topology.

   Students experiment with different network configurations that help then understand the various features of transparent bridges. In the second phase, they are asked to implement some of the features of the protocols.

6. *IP Fragmentation and Reassembly:* Students implement the fragmentation and reassembly protocol used in the Internet Protocol (IP). They experiment with six configurations, with varying MTU, link delays, and link errors. The purpose of this lab is to illustrate the behavior of a complex networking layer protocol in the presence of several link layers, each with a different MTU. In this way, the students understand the performance tradeoffs and overheads involved in fragmentation and reassembly.

7. *TCP Congestion Control:* Students implement the TCP window based flow control algorithm, the slow start alogrithm and congestion avoidance and fast retransmit/recovery algorithms. They study the performance of these algorithms for different error and congestion loss scenarios within the network. Other features of TCP like connection setup, and selective acknowledgments, can easily be added to the TCP model.

8. *ATM Traffic Management:* The goal of this lab is to understand the Generic Cell Rate Algorithm (GCRA) use by ATM networks for policing. The students implement the algorithm, and experiment with its behavior for a variety of traffic patterns.

9. *ATM PNNI Routing:* In this lab, students design, implement and experiment with a simplified version of the ATM PNNI routing protocol. They implement the Hello protocol, flooding of the PNNI Topology State Packets, and topology aggregation by the peer group leaders. This lab helps in the understanding of the hierarchical ATM PNNI addressing structure, and how each node learns the network topology. As the logical group nodes are created in the respective peer group leaders, the GUI shows a hierarchical formation of the logical nodes and the links associated with them.

# 5 Deployment Results

The labs developed so far have been used in the three courses on computer networking offered by the Department of Computer and Information Science at The Ohio State University. These courses are:

1. *CIS677: Introduction to Computer Networking.* The texts used in this course are [10] and [12]. This course is offered to upper level undergraduates, and covers the fundamental concepts and technologies in data communications and computer networking. Lab1 1–4 have been used in this class. Lab 5 may also be used depending on the material covered.

2. *CIS678: Internetworking (text [4]).* This course covers the topic of internetworking in detail, including

the various Internet Protocols, applications and security. Lab2 5–7 have been used in this course. Both undergraduates as well as graduate students take this course.

3. *CIS777: Telecommunications Networks (text [11])*. This course is primarily a graduate level course that covers recent developments in telecommunications networks. The topics covered are ISDN, B-ISDN, Frame Relay, SONET, and ATM. Labs 8–9 have been used for this class.

Another possibility is to use these labs in a laboratory course as part of a computer networking curriculum. This course would involve minimal lectures and would concentrate on lab projects. An advanced course could ask the students to implement new features for the existing protocol models, or even new protocol models for the environment.

The lab environment has brought about several changes in our networking curriculum.

First, it has provided a hands-on experience of networking concepts. Various methods of flow control, congestion control, error control, media access control, routing algorithms, and internetworking are topics that have traditionally been part of the networking curriculum [1, 5, 10, 12]. As discussed earlier, it is difficult for students to understand these topics without some visual experiences. The labs have enhanced the instruction and learning of these topics.

Second, these labs have combined recent developments in computer networking and telecommunications. The project combines the traditional packet switching topics with the newer cell switching approaches. ATM networks, which are the topic of labs 8 and 9, are a good example of protocols requiring knowledge of both fields. The labs have added newer telecommunications concepts of signaling and traffic management for Quality of Service traffic to the traditional networking concepts.

## 6 Concluding Remarks

In this paper, we have described a software package for a laboratory oriented curriculum in cmputer networking and telecommunications. The package consists of a simulation environment, together with a set of nine laboratory exercises covering a range of topics in networking. Tha package allows the student to experiment with existing protocol layer modules and architectures on a GUI, as well as to modify/develop modules using a C based programming language.

Currently, the software package is available for HP-UX 9.X, and Sun-OS environments. We are currently working the port the package for PCs using Linux as well as Windows 95. The package is available from the project site at http://www.netlab.ohio-state.edu/cise.

## 7 Acknowledgments

We would like to thank the National Science Foundation for supporting this project. We would also like to acknowledge the other (past and present) project members – Shivkumar Kalyanaraman, T. Lin, Shih-wei Lei, Xiangrong Cai and Jainping Jiang – for their contributions to the *CISE project*.